\documentstyle[12pt]{article}

\catcode`\@=11
\long\def\@makefntext#1{
\protect\noindent \hbox to 3.2pt {\hskip-.9pt
$^{{\ninerm\@thefnmark}}$\hfil}#1\hfill}		

 \def\@makefnmark{\hbox to 0pt{$^{\@thefnmark}$\hss}}  

\def\ps@myheadings{\let\@mkboth\@gobbletwo
\def\@oddhead{\hbox{}
\rightmark\hfil\ninerm\thepage}
\def\@oddfoot{}\def\@evenhead{\ninerm\thepage\hfil
\leftmark\hbox{}}\def\@evenfoot{}
\def\sectionmark##1{}\def\subsectionmark##1{}}

\newcounter{sectionc}\newcounter{subsectionc}\newcounter{subsubsectionc}
\renewcommand{\section}[1] {\vspace{0.6cm}\addtocounter{sectionc}{1}
\setcounter{subsectionc}{0}\setcounter{subsubsectionc}{0}\noindent
	{\bf\thesectionc. #1}\par\vspace{0.4cm}}
\renewcommand{\subsection}[1] {\vspace{0.6cm}\addtocounter{subsectionc}{1}
	\setcounter{subsubsectionc}{0}\noindent
	{\it\thesectionc.\thesubsectionc. #1}\par\vspace{0.4cm}}
\renewcommand{\subsubsection}[1]
{\vspace{0.6cm}\addtocounter{subsubsectionc}{1}
	\noindent {\rm\thesectionc.\thesubsectionc.\thesubsubsectionc.
	#1}\par\vspace{0.4cm}}

\newcounter{appendixc}
\newcounter{subappendixc}[appendixc]
\newcounter{subsubappendixc}[subappendixc]

\renewcommand{\appendix}[1] {\vspace{0.6cm}
        \refstepcounter{appendixc}
        \setcounter{figure}{0}
        \setcounter{table}{0}
        \setcounter{equation}{0}
        \renewcommand{\thefigure}{\Alph{appendixc}.\arabic{figure}}
        \renewcommand{\thetable}{\Alph{appendixc}.\arabic{table}}
        \renewcommand{\theappendixc}{\Alph{appendixc}}
        \renewcommand{\theequation}{\Alph{appendixc}.\arabic{equation}}
        \noindent{\bf Appendix \theappendixc #1}\par\vspace{0.4cm}}

\def\abstracts#1{{
	\centering{\begin{minipage}{30pc}\tenrm\baselineskip=12pt\noindent
	\centerline{\tenrm ABSTRACT}\vspace{0.3cm}
	\parindent=0pt #1
	\end{minipage}}\par}}

\renewenvironment{thebibliography}[1]
	{\begin{list}{\arabic{enumi}.}
	{\usecounter{enumi}\setlength{\parsep}{0pt}
\setlength{\leftmargin 1.25cm}{\rightmargin 0pt}
	 \setlength{\itemsep}{0pt} \settowidth
	{\labelwidth}{#1.}\sloppy}}{\end{list}}

\newcounter{itemlistc}
\newcounter{romanlistc}
\newcounter{alphlistc}
\newcounter{arabiclistc}

\newcommand{\fcaption}[1]{
        \refstepcounter{figure}
        \setbox\@tempboxa = \hbox{\tenrm Fig.~\thefigure. #1}
        \ifdim \wd\@tempboxa > 6in
           {\begin{center}
        \parbox{6in}{\tenrm\baselineskip=12pt Fig.~\thefigure. #1}
            \end{center}}
        \else
             {\begin{center}
             {\tenrm Fig.~\thefigure. #1}
              \end{center}}
        \fi}

\newcommand{\tcaption}[1]{
        \refstepcounter{table}
        \setbox\@tempboxa = \hbox{\tenrm Table~\thetable. #1}
        \ifdim \wd\@tempboxa > 6in
           {\begin{center}
        \parbox{6in}{\tenrm\baselineskip=12pt Table~\thetable. #1}
            \end{center}}
        \else
             {\begin{center}
             {\tenrm Table~\thetable. #1}
              \end{center}}
        \fi}

\def\@citex[#1]#2{\if@filesw\immediate\write\@auxout
	{\string\citation{#2}}\fi
\def\@citea{}\@cite{\@for\@citeb:=#2\do
	{\@citea\def\@citea{,}\@ifundefined
	{b@\@citeb}{{\bf ?}\@warning
	{Citation `\@citeb' on page \thepage \space undefined}}
	{\csname b@\@citeb\endcsname}}}{#1}}

\newif\if@cghi
\def\cite{\@cghitrue\@ifnextchar [{\@tempswatrue
	\@citex}{\@tempswafalse\@citex[]}}
\def\citelow{\@cghifalse\@ifnextchar [{\@tempswatrue
	\@citex}{\@tempswafalse\@citex[]}}
\def\@cite#1#2{{$\null^{#1}$\if@tempswa\typeout
	{IJCGA warning: optional citation argument
	ignored: `#2'} \fi}}

\def\fnt#1#2{\footnotetext{\kern-.3em
	{$^{\mbox{\sevenrm #1}}$}{#2}}}

 1
 1
 1

\font\tenbf=cmbx10
\font\tenrm=cmr10
\font\tenit=cmti10

\font\ninerm=cmr9

\begin{document}

\centerline{\tenbf SOME RESULTS ON $(0,2)$ ${\bf Z}_{N}$ ORBIFOLD}
\baselineskip=16pt
\centerline{\tenbf THEORIES WITH CONTINUOUS WILSON LINES\footnote{
Talk presented by T.M. at the Joint US Polish Workshop on
Physics from the Planck scale to the Electro--Weak scale,
Warsaw, 21.9. - 24.9.1994.}
}
\vspace{0.8cm}
\centerline{\tenrm GABRIEL LOPES CARDOSO, DIETER L{\"U}ST\footnote{
e-mail addresses:
GCARDOSO@QFT2.PHYSIK.HU-BERLIN.DE, LUEST@QFT1.PHYSIK.HU-BERLIN.DE}}
\baselineskip=13pt
\centerline{\tenit Humboldt Universit\"at zu Berlin}
\baselineskip=12pt
\centerline{\tenit Institut f\"ur Physik}
\baselineskip=12pt
\centerline{\tenit D-10115 Berlin, Germany}
\vspace{0.3cm}
\centerline{\tenrm and}
\vspace{0.3cm}
\centerline{\tenrm THOMAS MOHAUPT\footnote{
e-mail address: MOHAUPT@HADES.IFH.DE}
}
\baselineskip=13pt
\centerline{\tenit DESY-IfH Zeuthen}
\baselineskip=12pt
\centerline{\tenit Platanenallee 6}
\baselineskip=12pt
\centerline{\tenit D-15738 Zeuthen, Germany}
\vspace{0.9cm}

\abstracts{
We discuss recent results on orbifold compactifications with
(0,2) world sheet supersymmetry and continuous Wilson lines,
emphasizing the role of modular symmetries.
}

\vfil
\rm\baselineskip=14pt
\section{Introduction}

Orbifold compactifications \cite{DHVW1,DHVW2,NSV}
of the heterotic string are interesting
to study, because they are both exactly solvable and give rise
to a $N=1$ supersymmetric spectra with interesting gauge groups
and chiral matter. Whereas for a general compactification one
needs $(2,2)$ world sheet (WS) supersymmetry in order to compute
some quantities of interest, one can consider here models with
$(0,2)$ WS supersymmetry which is sufficent in order to have
$N=1$ space time (ST) supersymmetry.
If the model is constructed such that some
components of the Wilson line can still be varied continuously,
then the rank of the gauge group can be smaller then $16$
\cite{IbaNilQue} and its
level can be bigger then 1 \cite{FonIbaQue},
thus allowing reasonable gauge groups
and a realistic Higgs
sector for GUT models. The physical properties can be worked out
using the effective supergravity action valid below the Planck
or string scale. Here one uses the fact there is a one to one
correspondence between the moduli parametrizing deformations
of the string theory and scalars with flat potential in the effective
action. Thus the generalized kinetic term of these scalars is
fixed by the K\"ahler potential of the moduli metric.
The global structure of the moduli space is also reflected
by the effective action which must be invariant under the modular
group of the internal sector. This can be used \cite{FerLueShaThe}
once the K\"ahler
potential $K$ is known in order to restrict the form of the
superpotential $W$, because for example the effective potential
does only depend
on the combination $G=K+\log |W|^{2}$.

\rm\baselineskip=14pt
\section{Results}
\vspace{-0.7cm}

\subsection{Local structure of moduli space, K\"ahler potentials}
\vspace{-0.35cm}

The moduli of an orbifold model fall into two classes: the untwisted
and the twisted moduli.
The untwisted moduli are those moduli
of the original model, which are compatible with the twist and
therefore are still moduli of the twisted model. The twisted moduli,
which we will not study here, live in the the so called twisted
sectors that one has to add in order to preserve WS
modular invariance.

In the case of Narain compactifications the moduli space is locally
isomorhic to
$SO(22,6)$/$SO(22) \otimes SO(6)$ \cite{Nar}.  $SO(22,6)$ is
the group of allowed deformations of the Narain lattice
(which parametrizes the  moduli dependent part of the theory)
and the subgroup of Euclidean rotations is modded out, because
its action is physically trivial. If one now constructs
an asymmetric orbifold by modding out a rotation
$\Theta = \Theta_{L} \otimes \Theta_{R}$
$\in SO(22) \otimes SO(6)$, then the space of untwisted moduli
is locally given by the subgroup of $SO(22,6)$ commuting with
$\Theta$ modulo the subgroup of pure rotations. As shown in \cite{CLM}
the result is
\begin{equation}
{\cal M}_{O}(\Theta)
= \bigotimes_{i=1}^{n} \frac{SU(p_{i},q_{i})}{SU(p_{i}) \otimes
SU(q_{i}) \otimes U(1)}
\otimes \frac{SO(r,s)}{SO(r) \otimes
SO(s)}
\otimes \frac{SO(u,v)}{SO(u) \otimes
SO(v)}
\end{equation}
where $p_{i}, r, u$ and $q_{i}, s, v$ are the multiplicities
of the eigenvalues $e^{2\pi i \phi_{i}}, -1,1$ of
$\Theta_{L}$ and $\Theta_{R}$ respectively.
If $\Theta_{R}$ fulfills the constraint
$SU(2) \not\ni \Theta_{R} \in SU(3)$ imposed by $N=1$ ST
supersymmetry \cite{DHVW2}, then $v=0$ and either
$s=0$ or $s=2$ and the moduli space
is a complex K\"ahler manifold.
The K\"ahler potential can now be found along the lines of \cite{CLM,CLM2}.
In the limit of vanishing Wilson lines the results of
\cite{FerKouPor,CveLouOvr}
are reproduced.
Let us now look at the effect of continuous
Wilson lines in an concrete example and assume that
the internal torus factorizes as
${\bf T}^{6}={\bf T}^{2} \otimes {\bf T}^{4}$ such that the
internal twist acts as $-I_{2}$ on ${\bf T}^{2}$.

In this case there is a T and a U modulus as usual
\cite{DVV,CveLouOvr}, which are now supplemented by
two continous Wilson line associated with the
inequivalent directions on the torus, taking values
in that part of the Cartan subalgebra of the gauge group
on which the gauge twist acts as $-I$.
Assuming for defineteness
that this subspace has also dimension two, we have four real
Wilson moduli which can be rearranged into two complex ones,
$B$ and $C$. According to the general result the moduli space
is now
\begin{equation}
{\cal M}_{4,2} =   \frac{SO(4,2)}{SO(4) \otimes SO(2)}
\mbox{  and  }
K = - \log \left( (T+\overline{T})(U + \overline{U}) - \frac{1}{2}
(B + \overline{C})(C + \overline{B}) \right)
\end{equation}
is the K\"ahler potential as shown in \cite{CLM}.
For vanishing Wilson lines the moduli space factorizes
into two separate $SU(1,1)$/$U(1)$ cosets
corresponding to complex and to K\"ahler deformations.
Thus in this limit, where
the $(0,2)$ WS supersymmetry is enhanced to $(2,2)$
special geometry is restored.
The holomorphic mixing terms are physically interesting
because terms of this type can generate once local supersymmetry
is spontanously broken a Higgs mass term of order $m_{3/2}$
thus offering a solution to the $\mu$ problem \cite{GiuMas}.
As observed in \cite{KapLou,AntGavNarTay} such terms appear naturally
through matter fields in $(2,2)$ compactifications.

\subsection{Modular symmetries}
\vspace{-0.35cm}

Moduli spaces of string compactifications have a very non--trivial
global structure because some large deformations are actualy
automorphisms \cite{GivRabVen,DVV,Spa}.
This is analog to moduli spaces of
Riemann surfaces. The modular groups of $SO(r,2)$ and $SU(m,n)$ cosets
are denoted by $SO(r,2,{\bf Z})$ and $SU(m,n,{\bf Z})$, but
note that they depend on the details of the
underlying lattices and therefore some care is required in their
definition. For compactifications on
${\bf T}^{2}$/${\bf Z}_{2}$ with two complex Wilson line moduli
we find that the well known subgroup
$SL(2,{\bf Z})_{T}$ of the modular group acts by
\begin{equation}
T \rightarrow \frac{a T - i b}{icT + d},\;\;\;
U \rightarrow U - \frac{ic}{2} \frac{BC}{i c T +d},\;\;\;
B \rightarrow \frac{B}{icT + d},\;\;\;
C \rightarrow \frac{C}{icT + d}
\end{equation}
with $a,b,c,d \in {\bf Z},\;\;ad-bc = 1$.
Note that the same transformation law also applies to matter field
$(2,2)$ compactifications \cite{AntGavNarTay}.
As is evident from the so called mirror symmetry
$T \leftrightarrow U$,
there exists another subgroup
$SL(2,{\bf Z})_{U}$ with an obvious action on the moduli.

There are also modular transformations which are only
non--trivial when the Wilson lines are switched on. These contain
translations of the Wilson lines by lattice vectors \cite{Lau}
which have
the following action on the complex moduli \cite{CLM}:
\begin{equation}
T \rightarrow T+ p U + \frac{p}{2} \sqrt{2 }
(B+C),\;\;\;U\rightarrow U,\;\;\;
B \rightarrow B + p \sqrt{2}U,\;\;\;
C \rightarrow C + p \sqrt{2 }U,
\end{equation}
with $p \in {\bf Z}$.
All these transformations leave the K\"ahler potential invariant
up to a K\"ahler transformation.

\subsection{Modular Forms and Superpotentials}
\vspace{-0.35cm}

Finally we want to use the transformation properties of the
K\"ahler potential to construct some candidates for non--perturbative
superpotentials along the lines of \cite{FerKouLueZwi}.
The starting point is the mass formula for string states
in a ${\bf Z}_{2}$ sector of an $N=1$ orbifold, which we will
choose to be the one related to an $SO(4,2)$ coset and with
Wilson lines taking values in the CSA of a $SU(2)^{2}$
\cite{CLM2}:
\begin{equation}
\alpha' M^{2} = \frac{|{\cal M}|^{2}}{Y} + 2 (N-1 + l^{T}l + m^{T}n)
\end{equation}
The first term on the right hand side is moduli dependent
and contains the holomorpic mass ${\cal M}$ \cite{FerKouLueZwi}
and a real analytic
function $Y$ which is related to the K\"ahler potential by
$K = - \log Y$. The second term is moduli independent and
contains the number $N$ of left moving excitations together
with the vectors $l=(l_{i})$, $n = (n_{i})$ and $m = (m_{i})$,
$i=1,2$
of charge, winding and momentum quantum numbers. The term
$l^{T}l + m^{T}n$ is proportional to the $SO(4,2)$ invariant
scalar product.
Setting the second term to zero gives an $SO(4,2)$
invariant constraint which restricts the quantum numbers to
live in a certain orbit ${\cal O}$ of $SO(4,2,{\bf Z})$. Comparing
the mass formula to the form of the $G$ function,
one learns that $W_{\cal O}$, where
$\log W_{\cal O}  = \sum_{\cal O} \log {\cal M}$,
is a candidate for a non--perturbative superpotential
because it transforms formally in the right way.
Since the sums in general diverge some regularization
is needed.
Two interesting orbits are given by
${\cal O}_{1}:$
$N = 1 \Rightarrow $ $ l^{T}l + m^{T}n = 0$
and
${\cal O}_{0}:$
$N= 0 \Rightarrow  $$l^{T}l + m^{T}n =1$.
Here, we will only consider the following simplest $SO(2,2,{\bf Z})$-
invariant suborbits, namely
${\cal O}_{1,0}$: $l=0, m^{T}n=0$ and
${\cal O}_{0,1},\ldots {\cal O}_{0,4}$: $l^{T}l = 1$, $m=n=0$,
which give rise to three inequivalent functions,
\begin{equation}
W_{1} = \eta^{-2}(U) \eta^{-2}(T) \left[
1 - \frac{BC}{2} \partial_{U} \log \eta^{2}(U)
\partial_{T} \log \eta^{2}(T) + O( B^{2}C^{2} )
\right],
\label{BC}
\end{equation}
$W_{2} \simeq B+C$ and $W_{3} \simeq C-B$.
As expected these functions are (in case of $W_{1}$ to leading
order in $BC$) only covariant with respect to
$SO(2,2,{\bf Z})$ but fail to transform correctly under
shifts of the Wilson lines.
Note that $W_{1}$ is the
same superpotential as was postulated in \cite{AntGavNarTay}
in the context of $(2,2)$ orbifolds because
of its correct transformation properties under  $SO(2,2,{\bf Z})$.
We will give a more complete discussion of these issues in
\cite{CLM2}.
Note that the $BC$-terms in (\ref{BC}) yield an alternative
solution to the the $\mu$- problem \cite{CasMun}.

\rm


\baselineskip=14pt
\section{References}
\vspace{-0.6cm}





\begin{thebibliography}{9}
\bibitem{DHVW1}
        L. Dixon, J. Harvey, C. Vafa, and E. Witten,
        {\em Nucl. Phys.} {\bf B 261}
        (1985)
        678.
\bibitem{DHVW2}
        L. Dixon, J. Harvey, C. Vafa, and E. Witten,
        {\em Nucl. Phys. } {\bf B 274}
        (1986)
        285.
\bibitem{NSV}
        K. Narain, M. Sarmadi and C. Vafa,
        {\em Nucl. Phys.} {\bf B 288} (1987) 551.
\bibitem{IbaNilQue}
         L. E. Ib\'{a}\~{n}ez, H. P. Nilles and F. Quevedo,
         {\em Phys Lett.} {\bf B 192} (1987) 332.
\bibitem{FonIbaQue}
         A. Font, L. E. Ib\'{a}\~{n}ez and F. Quevedo,
         {\em Nucl. Phys.} {\bf B 345} (1990) 389.
\bibitem{FerLueShaThe}
         S. Ferrara, D. L\"ust, A. Shapere and S. Theisen,
         {\em Phys. Lett.} {\bf 255} (1989) 363.
\bibitem{Nar}
        K. Narain,
        {\em Phys. Lett.} {\bf B 169}
        (1986)
        41.
\bibitem{CLM}
        G. Lopes Cardoso, D. L{\"u}st and T. Mohaupt,
        hep-th-9405002.
\bibitem{CLM2}
        G. Lopes Cardoso, D. L{\"u}st and T. Mohaupt,
        in preparation.
\bibitem{FerKouPor}
        S. Ferrara, C. Kounnas and M. Porrati,
        {\em Phys. Lett.} {\bf B 181} (1986) 263.
\bibitem{CveLouOvr}
      M. Cvetic, J. Louis and B. A. Ovrut,
      {\em Phys. Lett. B} {\bf 206} (1988) 227.
\bibitem{DVV}
        R. Dijkgraaf, E. Verlinde, H. Verlinde,
        THU-87/30.
\bibitem{GiuMas}
        G.F. Giudice and A. Masiero, {\em Phys. Lett.} {\bf B 206}
        (1988) 480.
\bibitem{KapLou}
        J. Louis and V. Kaplunovsky, {\em Phys. Lett.} {\bf B 306}
        (1993) 269.
\bibitem{AntGavNarTay}
        I. Antoniadis, E. Gava, K.S. Narain and T.R. Taylor,
        hep-th 9405024.
\bibitem{GivRabVen}
        A. Giveon, E. Rabinovici, and G. Veneziano,
        {\em Nucl. Phys.} {\bf B 322}
        (1989)
        167.
\bibitem{Spa}
      M. Spalinski,
      {\em Phys. Lett.} {\bf B 275} (1992) 47.
\bibitem{Lau}
        J. Lauer, unpublished.
\bibitem{FerKouLueZwi}
        S. Ferrara, C. Kounnas, D. L\"ust, F. Zwirner,
        {\em Nucl. Phys.} {\bf B 365} (1991) 431.
\bibitem{CasMun}
        J. A. Casas and C. Munoz,
        {\em Phys. Lett.} {\bf B 306} (1993) 288.


\end{thebibliography}
\end{document}